\documentclass[aps,prb,amsmath,amssymb,reprint,superscriptaddress,preprintnumbers,showpacs,intlimits]{revtex4-1}
\usepackage{bm,latexsym,mathrsfs,enumerate,color}
\usepackage[mathcal]{euscript}
\usepackage{hyperref}
\usepackage{graphicx}
\usepackage{subcaption}

\renewcommand{\vec}[1]{\bm{#1}}
\begin{document}

\title{Saturation of magnetic films with spin-polarized current in presence of magnetic field}

\author{Oleksii M. Volkov}
\email{alexey@volkov.ca}
\affiliation{Taras Shevchenko National University of Kyiv, 01601 Kyiv, Ukraine}

\author{Volodymyr P. Kravchuk}
 \affiliation{Bogolyubov Institute for Theoretical Physics, 03143 Kyiv, Ukraine}

\date{December 14, 2012}

%
%

\begin{abstract}
Influence of perpendicular magnetic field on the process of transversal saturation of ferromagnetic films with spin-polarized current is studied theoretically. It is shown that the saturation current $J_s$ is decreased (increased) in case of codirected (oppositely directed) magnetic field and current. There exists a critical current $J_c>J_s$ which provides "rigid" saturation -- the saturated state is stable with respect to the transverse magnetic field of any amplitude and direction. Influence of the magnetic field on the vortex-antivortex crystals, which appear in pre-saturated regime, is studied numerically. All analytical results are verified using micromagnetic simulations.
\end{abstract}




\maketitle

	\section{Introduction}
	\label{sec:intro}
The influence of spin-polarized current on planar magnetic systems is of high applied and academic interest now. It is so mainly due to the possibility to handle the magnetization states of magnetic nanoparticles (nanomagnets) without using the external magnetic fields of complex space-time configurations. That provides new opportunities in construction of purely current controlled devices\cite{Lindner10}, e.g. magnetic disk drivers or Magnetic Random Access Memory (MRAM)\cite{Bohlens08,Drews09}.
	
\begin{figure}
\includegraphics[width=0.7\columnwidth]{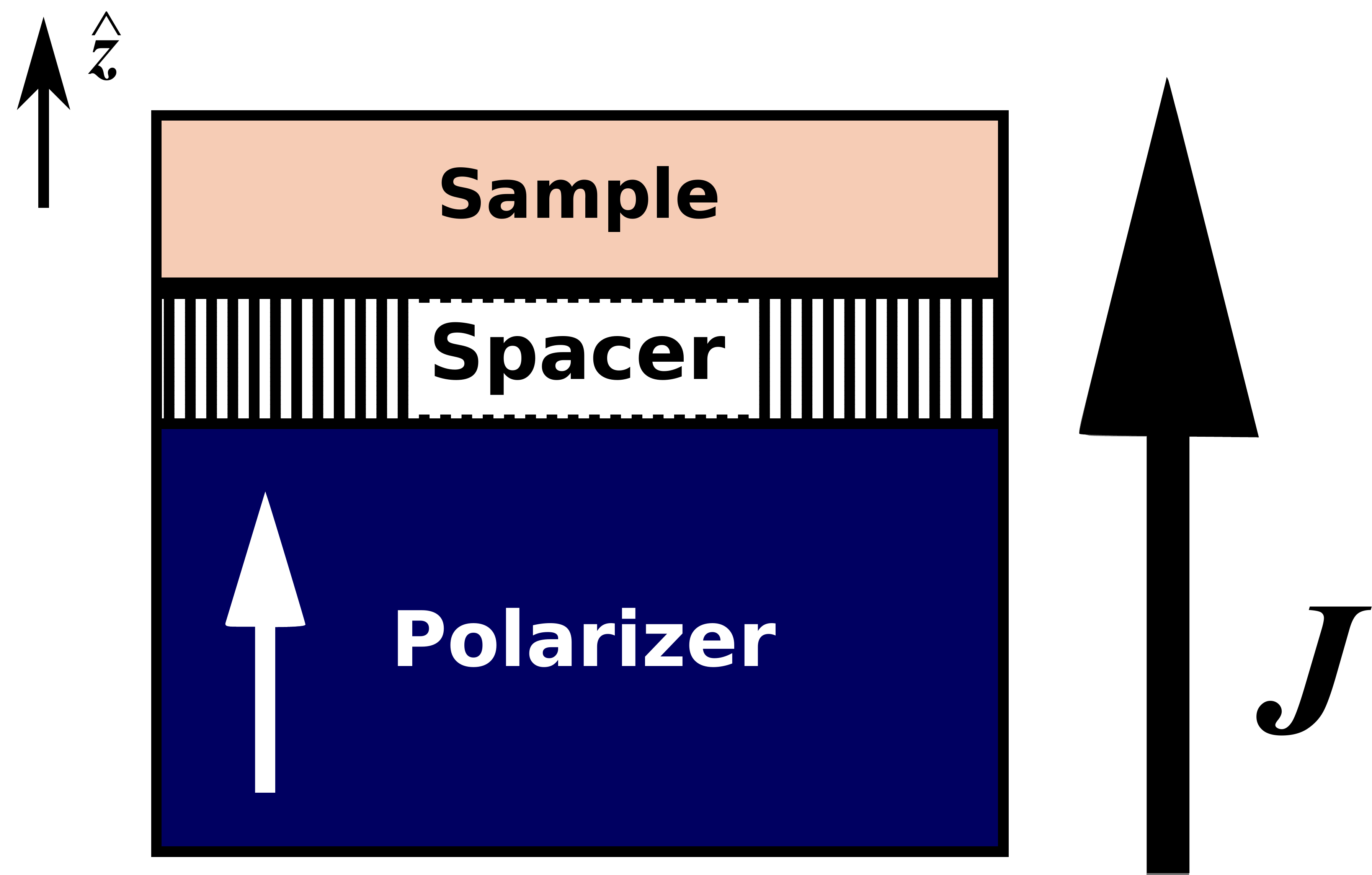}
\caption{The simplest pillar heterostructure consists of two ferromagnetic layers: Polarizer and Sample, and one nonmagnetic layer between them (Spacer). Black (larger) arrow shows the current direction, which flow through the heterostructure, and white (smaller) arrow shows the magnetization direction of the Polarizer.}\label{fig:Heter}
\end{figure}

A convenient way to provide the influence of spin-polarized current on the magnetic film is to use a pillar structure which was firstly proposed in Ref.~\cite{Kent04}. The simplest pilar structure consists of two ferromagnetic layers (Polarizer and Sample) and nonmagnetic Spacer between them, see Fig.~\ref{fig:Heter}. When the electrical current passes through the Polarizer the conduction electrons become partially spin-polarized in direction which is determined by the Polarizer magnetization. Polarizer is usually made of a hard ferromagnetic material whose magnetization is kept fixed. Spacer, being very thin (few nanometers), does not change spin polarization of the current electrons but it prevents the exchange interaction between Polarizer and Sample. Thus the spin-polarized electrons transfer the spin-torque from Polarizer to the Sample what can result in dynamics of the Sample magnetization.  The spin-torque influence can be described phenomenologically by  adding the Slonczewski-Berger term into Landau-Lifshitz equation \cite{Slonczewski96,Berger96,Slonczewski02}.

Recently we studied influence of strong spin-current on the magnetic films\cite{Volkov11, Gaididei12a}. It was shown that the strong spin-polarized current can saturate magnetic film and value of the saturation current density $J_s$ increases with the film thickness increasing. We also demonstrated that in the pre-saturated regime a stable vortex-antivortex lattices (VAL) appear. As it was recently shown\cite{Dussaux10,Dussaux12} the external magnetic field can drastically modify the magnetic system dynamics induced by the spin-torque. The aim of this paper is to study the influence of perpendicular magnetic field on the process of the film saturation with spin-current. For this purpose, we modify developed in Ref.~\cite{Gaididei12a} linear theory of instability of the saturated state for the case of presence of magnetic field and uniaxial anisotropy. It enable us to obtain the dependence of saturation current $J_s$ on the field amplitude. We also demonstrate that in linear approximation the actions of the perpendicular magnetic field and uniaxial anisotropy on the stability of saturated state are equivalent.  Using micromagnetic simulations we study how the properties of the VAL, which appear it the pre-saturated regime, depend on the value of the applied field.
	
	\section{The theory of saturated state stability}
	\label{sec:model}

Let us consider a ferromagnetic film with thickness $h$ and lateral size $L\gg h$. We use here a discrete model of the magnetic media considering a three-dimensional cubic lattice of magnetic moments $\vec{M_\nu}$ with lattice spacing $a \ll h$, where $\vec{\nu}=a (\nu_x,\nu_y,\nu_z)$ is a three-dimensional index with $\nu_x,\nu_y,\nu_z\in\mathbb{Z}$ (here and below all Greek indexes are three-dimensional and Latin indexes are two-dimensional). We assume also that $h$ is small enough to ensure the magnetization uniformity along the thickness. In this case one can use the two-dimensional discrete Landau-Lifshitz-Slonczewski equation \cite{Slonczewski96,Berger96,Slonczewski02}:
	\begin{equation} \label{eq:LLS}
		\dot{\vec{m}}_{\vec n} = \vec m_{\vec n}\times\dfrac{\partial\mathcal{E}}{\partial\vec{m}_{\vec n}}-\varkappa \dfrac{\vec m_{\vec n}\times[\vec m_{\vec n}\times\hat{\vec z}]}{1+(\vec m_{\vec n} \cdot \vec{\hat z})} ,
	\end{equation}
to describe the magnetization dynamics under the influence of a spin-polarized current which flows perpendicularly to the magnetic plane, along the $\hat{\vec z}$-axis, see Fig.~\ref{fig:Heter}. It is also assumed that the current flow and its spin-polarization are of the same direction in Eq.~\eqref{eq:LLS}. The two-dimensional index $\vec n= a (n_x, n_y)$ with $n_x,n_y\in\mathbb{Z}$ numerates the normalized magnetic moments $\vec m_{\vec n} = \vec M_{\vec n}/|\vec M_{\vec n}|$ within the film plane. The overdot indicates derivative with respect to the rescaled time in units of $(4\pi\gamma M_s)^{-1}$, $\gamma$ is gyromagnetic ratio, $M_s$ is the saturation magnetization, and $\mathcal{E}=E/(4\pi M_s^2 a^3 \mathcal{N}_z)$ is dimensionless magnetic energy, where $\mathcal{N}_z=h/a$ is the number of magnetic moments along the thickness. The normalized current density is presented by parameter $\varkappa=\eta J/J_0$ , where $\eta$ is the degree of spin polarization, $J$ is the current density, and $J_0=4\pi M_s^2|e|h/\hbar$ with $e$ being electron charge and $\hbar$ being Planck constant.

The Eq.~\eqref{eq:LLS} is written for the case when the conductance of the Sample is much lower than the conductance of the Spacer, what corresponds to high level of spin accumulation at the nonmagnet–ferromagnet interfaces.
The mismatch between spacer and ferromagnet resistances is traditionally described by $\Lambda$-parameter\cite{Slonczewski02,Sluka11}. But as it was shown in the Ref.~\cite{Gaididei12a} parameter $\Lambda$ is not included in the linearized problem and therefore it has no influence on the saturation process, that is why we do not include $\Lambda$ into our model assuming $\Lambda\gg1$. We also omitted damping in the equation of motion \eqref{eq:LLS}, because, as it was shown earlier\cite{Gaididei12a}, the spin-current provides an effective damping which is much larger than natural damping.
	
We consider here a magnetic system, the total energy $E=E_\mathrm{ex}+E_\mathrm{d}+E_\mathrm{z}+E_\mathrm{an}$ of which consists of four parts: exchange, dipole-dipole, Zeeman and magnetocrystalline anisotropy contributions. Exchange energy up to a constant has the form
	\begin{equation} \label{eq:Eex}
		E_\mathrm{ex}=-\mathcal{S}^2\mathcal{N}_z\mathcal{J}_{0}\sum\limits_{\vec n,\tilde{\vec n}}\vec m_{\vec n}\cdot\vec m_{\vec n+\tilde{\vec n}},
	\end{equation}
where $\tilde{\vec n}$ numerates the nearest neighbors within the film plane of $\vec n$-th atom, $\mathcal{S}$ is value of spin of a ferromagnetic atom, and $\mathcal{J}_{0}>0$ denotes the exchange integral between nearest atoms.
	
	The energy of dipole-dipole interaction is
	\begin{equation} \label{eq:Ems}
			E_\mathrm{d}=\frac{M_s^2a^6}{2}\!\sum\limits_{\vec\nu\ne\vec\lambda}\biggl[\frac{ (\vec m_{\vec\nu}\!\cdot\! \vec m_{\vec\lambda})}{r_{\vec\lambda\vec\nu}^3}
			-3\frac{\left(\vec m_{\vec\nu}\!\cdot\! \vec r_{\vec\lambda\vec\nu}\right) \left(\vec m_{\vec\lambda}\!\cdot\! \vec r_{\vec\lambda\vec\nu}\right)}{r_{\vec\lambda\vec\nu}^5}\biggr],
	\end{equation}
where $\vec r_{\vec\lambda\vec\nu}=\vec\lambda-\vec\nu$ with $\vec\lambda$ and $\vec\nu$ being the three-dimensional indexes.
	
	The Zeeman energy describes the interaction of magnetic film with external perpendicular magnetic field $\vec{B}= B \vec{\hat{z}}$ and it reads
	\begin{equation} \label{eq:Ez}
		E_\mathrm{z}= -B M_s a^3 \mathcal{N}_z  \sum\limits_{\vec{n}} m^z_{\vec{n}}.
	\end{equation}		
	
And finally we introduce the energy of uniaxial anisotropy, which axis is oriented perpendicularly to the film plane:
	\begin{equation} \label{eq:Ean}
		E_\mathrm{an}= -\frac{K}{2} a^3 \mathcal{N}_z \sum\limits_{\vec{n}}(m^z_{\vec{n}})^2	
	\end{equation}		
where $K$ is the anisotropy coefficient which can be positive (easy-axis) as well as negative (easy-plane) value.	
	
	By introducing the complex variable \cite{Gaididei12a}
	\begin{equation} \label{eq:psi}
		\psi_{\vec n}=\frac{m^x_{\vec n}+im^y_{\vec n}}{\sqrt{1+m^z_{\vec n}}},
	\end{equation}
one can write the Eq. \eqref{eq:LLS} in form
	\begin{equation} \label{eq:LLS-psi}
	i\dot\psi_{\vec n}=-\frac{\partial\mathcal{E}}{\partial\psi_{\vec n}^*}-i\frac{\partial\mathcal{F}}{\partial\psi_{\vec n}^*},
	\end{equation}
where function
\begin{equation}
\mathcal{F}=\frac{\varkappa}{2}\sum\limits_{\vec n}|\psi_{\vec n}|^2
\end{equation}
represents an action of the spin-polarized current.

	For the future analysis it is convenient to proceed to the wave-vector representation using the two-dimensional discrete Fourier transform
	\begin{subequations}\label{eq:Fourier-def}
		\begin{align}
			\label{eq:four-inv}&\psi_{\vec n}=\frac{1}{\sqrt{\mathcal{N}_{xy}}}\sum\limits_{{\vec k}}\hat\psi_{\vec k}e^{i \vec k\cdot\vec n},\\
			\label{eq:four}&\hat\psi_{\vec k}=\frac{1}{\sqrt{\mathcal{N}_{xy}}}\sum\limits_{{\vec n}}\psi_{\vec n}e^{-i \vec k\cdot\vec n}
		\end{align}
	\end{subequations}
with the orthogonality condition
	\begin{equation} \label{eq:orth-cond}
		\sum\limits_{{\vec n}}e^{i(\vec k-\vec k')\cdot\vec n}=\mathcal{N}_{xy}\Delta(\vec k-\vec k'),
	\end{equation}
where $\mathcal{N}_{xy}=L^2/a^2$ is the total number of atoms within the film plane, $\vec k=(k_x,k_y)\equiv\frac{2\pi}{L}(l_x, l_y)$ is two-dimensional discrete wave vector, $l_x, l_y\in\mathbb{Z}$, and $\Delta(\vec k)$ is the Kronecker delta.

Applying \eqref{eq:Fourier-def} to the equation \eqref{eq:LLS-psi} one obtains equation of motion in reciprocal space:
	\begin{equation} \label{eq:main-Four}
		-i\dot{\hat{\psi}}_{\vec k}=\frac{\partial\mathcal{E}}{\partial\hat\psi_{\vec k}^*}+i\frac{\partial\mathcal{F}}{\partial\hat\psi_{\vec k}^*},
	\end{equation}

Since we are studying the stability of the saturated state we can linearize the equation of motion \eqref{eq:main-Four} in vicinity of the solution $m^z_{\vec n}=1$ what is equivalent to $|\psi_{\vec n}|=0$ and $|\hat\psi_{\vec k}|=0$. To obtain the energy functional $\mathcal{E}$ in ``$\psi$''-representation we expand components of the magnetization vector into series in the way similar to the representation in terms of the Bose operators \cite{Akhiezer68}:
	\begin{equation} \label{eq:mxmymz}
		\begin{split}
			&m^x_{\vec n}=\frac{\psi_{\vec n}+\psi_{\vec n}^*}{\sqrt{2}}+\mathcal{O}(|\psi_{\vec n}|^3),\\
			&m^y_{\vec n}=\frac{\psi_{\vec n}-\psi_{\vec n}^*}{i\sqrt{2}}+\mathcal{O}(|\psi_{\vec n}|^3),\\
			&m^z_{\vec n}=1-|\psi_{\vec n}|^2.
		\end{split}
	\end{equation}

Substituting \eqref{eq:mxmymz} into energy components \eqref{eq:Eex}, \eqref{eq:Ems}, \eqref{eq:Ez}, \eqref{eq:Ean} and applying the Fourier transform \eqref{eq:Fourier-def}, \eqref{eq:orth-cond} one can write the dimensionless energy functional in form
\begin{subequations}\label{eq:harmonic}
\begin{align} \label{eq:all-energies}
\mathcal{E}=\mathcal{E}_\mathrm{ex}+\mathcal{E}_\mathrm{d}+
\mathcal{E}_\mathrm{z}+\mathcal{E}_\mathrm{an}.
\end{align}
Here the exchange contribution reads \cite{Gaididei12a}
\begin{align} \label{eq:Eex-lin}
\mathcal{E}_\mathrm{ex}=\ell^2\sum\limits_{\vec k}|\hat\psi_{\vec k}|^2k^2 +\mathcal{O}(|\hat\psi_{\vec k}|^4),
\end{align}
where $\ell=\sqrt{\mathcal{S}^2\mathcal{J}_{0}c/(4\pi M_s^2a)}$ is the exchange length with $c=4$ being the number of nearest neighbors within the film plane.	
The energy of dipole-dipole interaction has the form \cite{Gaididei12a}
\begin{equation} \label{eq:Ed-lin}
	\begin{split}
		&\mathcal{E}_\mathrm{d} =\sum\limits_{\vec k}\left[\frac{g(kh)}{2}-1\right]|\hat\psi_{\vec k}|^2\\
		&+\frac{g(kh)}{4}\left[\frac{(k^x-ik^y)^2}{k^2}\hat\psi_{\vec k}\hat\psi_{-\vec k}+\text{c.c.}\right] +\mathcal{O}(|\hat\psi_{\vec k}|^4),
	\end{split}
\end{equation}
where $g(x)\equiv(x+e^{-x}-1)/x$.
The Zeeman energy takes the form
	\begin{equation} \label{eq:Ez-lin}
		\mathcal{E}_\mathrm{z}= \beta\sum\limits_{k}|\hat\psi_{\vec k}|^2,
	\end{equation}
where $\beta=B/(4 \pi M_s )$ is the dimensionless magnetic field in units of the saturation field.
And anisotropy energy can be written as
	\begin{equation}\label{eq:Ean-lin}
		\mathcal{E}_\mathrm{an}= \alpha \sum\limits_{k}|\hat\psi_{\vec k}|^2+\mathcal{O}(|\hat\psi_{\vec k}|^4).
	\end{equation}	
where $\alpha=K/(4 \pi M_s^2 )$ is the dimensionless anisotropy coefficient.	
\end{subequations}

Details of deriving the contributions \ref{eq:Eex-lin} and \ref{eq:Ed-lin} in the wave-vector space can be found in Appendix A of the Ref.~\cite{Gaididei12a}.

The current-action function $\mathcal{F}$ in the wave space has the simple form
current:
	\begin{equation} \label{eq:F-function}
		\mathcal{F}=\frac{\varkappa}{2}\sum\limits_{\vec k} |\hat\psi_{\vec k}|^2.
	\end{equation}

\begin{figure}
\includegraphics[width=\columnwidth]{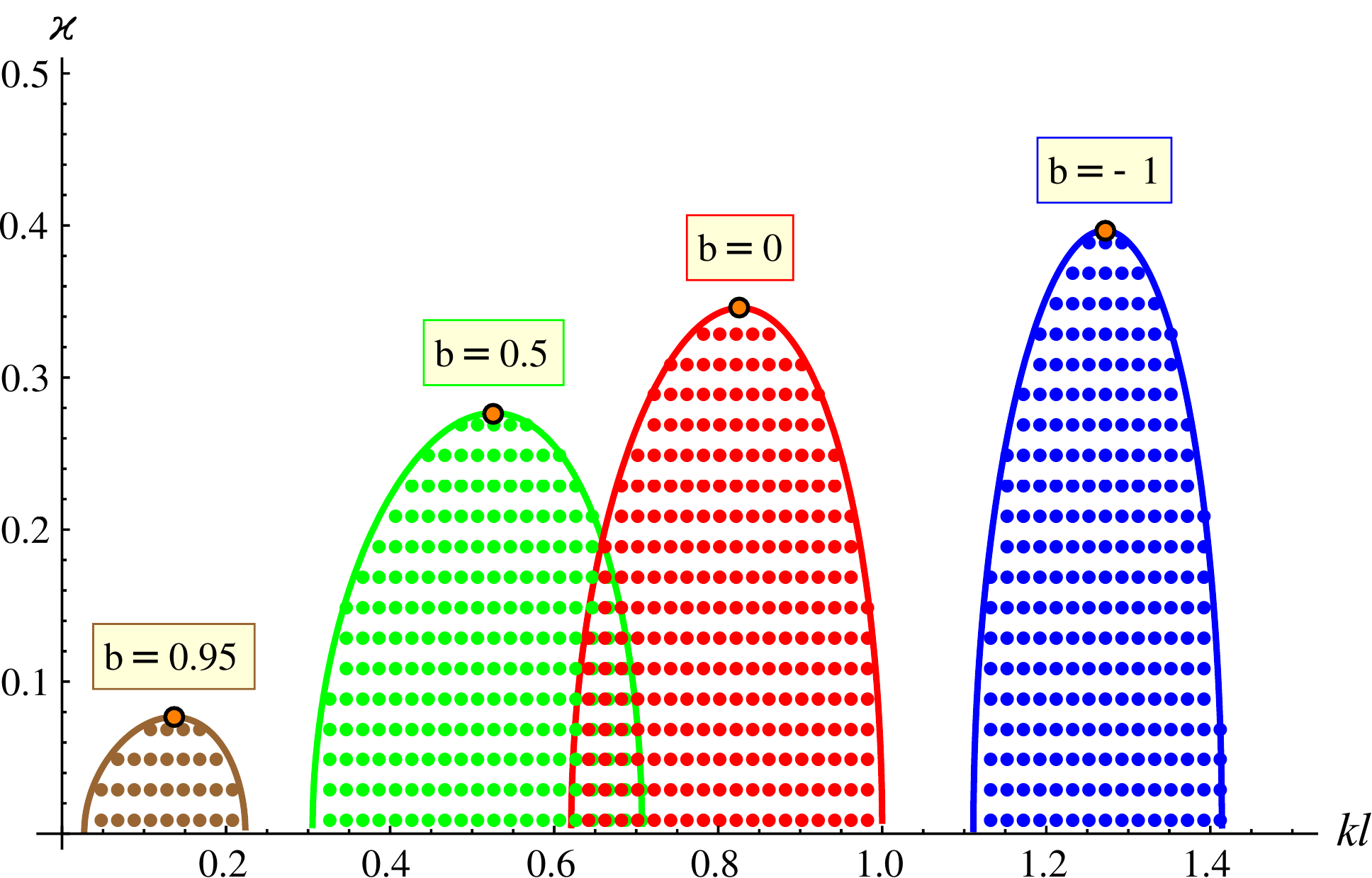}
\caption{(Color online) The influence of parameter $b$ on the instability domains (filled regions) of the transversally saturated Permalloy ($\ell=5.3$ nm) film with thickness $h=20$ nm. The instability domains are determined by condition \eqref{eq:condition}. Points at the maximums determine the saturation current $\varkappa_s$ for the given $b$.}\label{fig:stab-diagramm}
\end{figure}

Substituting \eqref{eq:harmonic} and \eqref{eq:F-function} into \eqref{eq:main-Four} one obtains the system of linear equation for the complex amplitudes $\hat\psi_{\vec k}$ and $\hat\psi_{-\vec k}^*$:
	\begin{equation} \label{eq:eq-of-motion}
		\begin{split}
			-i\dot{\hat{\psi}}_{\vec k}   &=  \left[k^2\ell^2-1+\frac{g(hk)}{2}+b+i\dfrac{\varkappa}{2}\right]\hat\psi_{\vec k}\\
			&+\frac{g(hk)}{2}\frac{(k^x+ik^y)^2}{k^2}\hat{\psi}^*_{-\vec k},\\
			i\dot{\hat{\psi}}_{-\vec k}^* &= \left[k^2\ell^2-1+\frac{g(hk)}{2}+b-i\dfrac{\varkappa}{2}\right]\hat\psi_{-\vec k}^*\\
			&+\frac{g(hk)}{2}\frac{(k^x-ik^y)^2}{k^2}\hat{\psi}_{\vec k}.
\end{split}
	\end{equation}
where $b=\alpha+\beta$.

\begin{figure}
\includegraphics[width=\columnwidth]{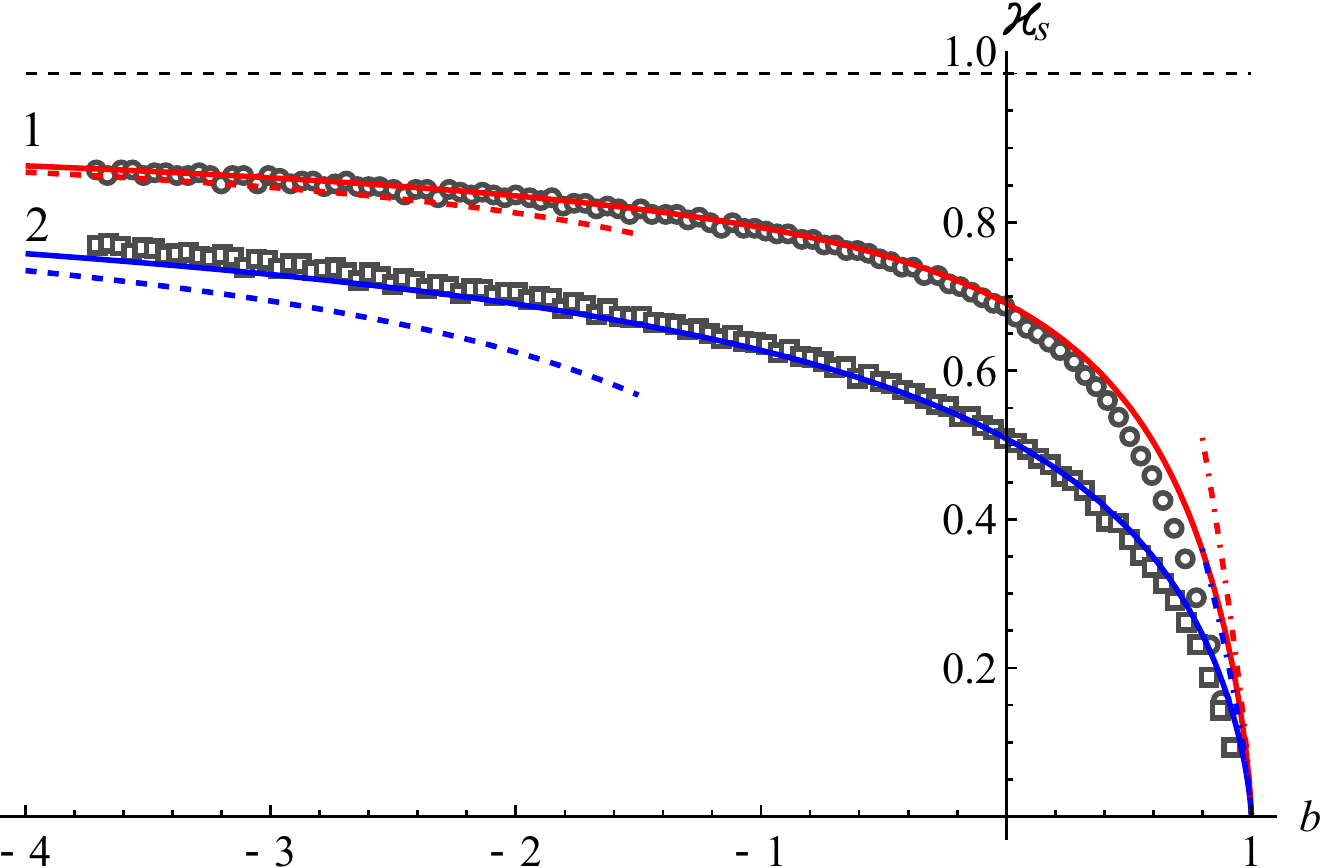}
\caption{(Color online) Dependence of the saturation current $\varkappa_c$ on the magnetic field. Solid line corresponds to the analytical solution obtained from \eqref{eq:kappa} and results of micromagnetic simulations (see text) are shown by markers. Dashed lines correspond to asymptotic \eqref{eq:asympt-b-} and dot-dashed lines show the asymptotic \eqref{eq:asympt-b1}. Dependencies 1 and 2 corresponds to different thicknesses: $h=20$ nm and $h=10$ nm respectively.}\label{fig:B_Jc}
\end{figure}

Looking for solutions of Eq.~\eqref{eq:eq-of-motion} in the form
\begin{equation} \label{eq:psi-sol}
\hat{\psi}_{\vec k}=\Psi_{+}e^{z(k)t},\qquad \hat{\psi}_{-\vec k}^*=\Psi_{-}e^{z(k)t},
\end{equation}
where $\Psi_{\pm}$ are time independent amplitudes, one obtains the following condition for the rate constant $z(k)$
\begin{equation} \label{eq:dispersion}
z(k)=-\frac{\varkappa}{2}\pm\tilde\varkappa(k).
\end{equation}
Here the function $\tilde\varkappa(k)$ is given by
\begin{subequations}\label{eq:kappa}
	\begin{equation} \label{eq:crit-current}
		\tilde\varkappa(k) =\sqrt{\left(1-k^2\ell^2-b \right)\left( k^2\ell^2+g(hk)-1+b \right)}.
	\end{equation}

	Accordingly to the Eq.~\eqref{eq:dispersion} one can conclude that if value of the function $\tilde{\varkappa}(k)$ is complex then the saturated state of the film is stable. If value of the function $\tilde{\varkappa}(k)$ is real then we have two different cases: for strong currents when $\varkappa > 2\tilde{\varkappa}$ we have $\mathrm{Re}\,z(k)<0$ and therefore the stationary state of the system is the saturated state with $m_z=1$. However, for smaller currents $\varkappa < 2\tilde{\varkappa}$ the instability of the saturated state develops. Functions $2\tilde{\varkappa}(k)$ for different values of parameter $b$ is shown in the Fig.~\ref{fig:stab-diagramm}. One can see that $\tilde{\varkappa}(k)$ is a  nonmonotonic function, which reaches its maximum value $\tilde\varkappa_c$ at $k=K$:
\begin{equation}\label{eq:crit-current-max}
\frac{\mathrm{d }\tilde\varkappa(k)}{\mathrm{d} k}=0, \quad \tilde\varkappa_c=\max\limits_k\tilde\varkappa(k)\equiv\tilde\varkappa(K_0).
\end{equation}
Thus the value $\tilde\varkappa_c$ determines the lowest current at which the saturated state remains stable -- the saturation current:
\begin{equation}\label{eq:kappa-s}
\varkappa_s=2\tilde\varkappa_c\Rightarrow J_s=2\frac{J_0}{\eta} \tilde\varkappa_c.
\end{equation}
\end{subequations}
The instability domains, which are determined by condition
\begin{equation}\label{eq:condition}
\tilde\varkappa\in\mathbb{R},\quad\text{and}\quad\varkappa<\varkappa_s,
\end{equation}
are shown in the Fig.~\ref{fig:stab-diagramm} as filled regions. Maximum values of the shown dependencies determine the saturation current $\varkappa_s$ for the given value of the $b$-parameter. Dependence $\varkappa_s(b)$ is shown in the Fig.~\ref{fig:B_Jc}
One can see that for $b>1$ the magnetic film is perpendicularly saturated without current. For example, this case can be realized for magnetically soft film ($\alpha=0$) when the external field exceed the saturation value $\beta>1$. Analysis of \eqref{eq:kappa} enable one to obtain the following asymptotical behaviors
\begin{subequations}\label{eq:asympt}
\begin{equation}\label{eq:asympt-b1}
\varkappa_s\approx2\sqrt\frac{h}{\ell}\left(\frac{1-b}{3}\right)^{3/4},\qquad b\lessapprox1,
\end{equation}

\begin{figure*}
\includegraphics[width=0.85\textwidth]{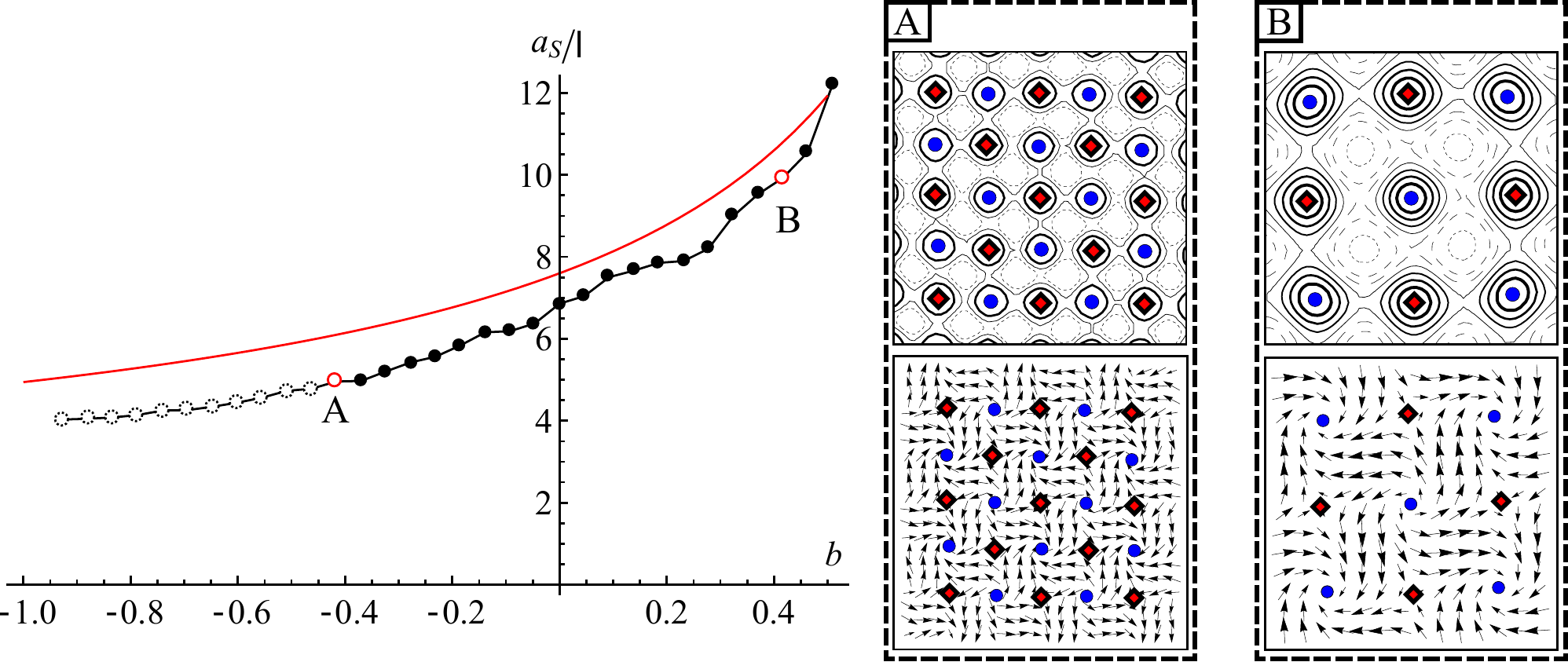}
\caption{(Color online) Dependence of period of VAL, which appears in the pre-saturated regime, on the applied field. The data obtained using micromagnetic modelling (see test) are shown by markers, dark disks corresponds to the VAL, whereas the dashed circles represents the fluid-like state of the vortex-antivortex system (see Ref.~\cite{Volkov11},\cite{Gaididei12a}). The solid line corresponds to dependence $2\pi/K_0$ obtained from \eqref{eq:crit-current-max}. The insets A and B show the magnetization distribution in VAL for cases of positive and negative fields respectively. The upper parts of the insets show the $m_z$-component distribution: with decreasing the value of $m_z$ the contour line becomes thinner and when $m_z$ is close to the minimum value the contours become dashed. Magnetization distribution within the film plane is shown by arrows in the bottom parts of the insets, disks and diamonds show centers of vortices and antivortices respectively. The size of the inset area is $50\times50$ nm. }\label{fig:as_vs_b}
\end{figure*}

\begin{equation}\label{eq:asympt-b-}
\varkappa_s\approx1-\frac{\ell}{h\sqrt{|b|}},\qquad b\rightarrow -\infty.
\end{equation}
\end{subequations}
Accordingly to \eqref{eq:asympt-b-} the saturation current is a bounded above quantity: $\varkappa_s<1$ for any values of parameters. It means that for currents $\varkappa\geqslant1$ the perpendicularly saturated magnetic film remains stable for any values of magnetic field and uniaxial anisotropy constant. In other words, if the current $\varkappa\geqslant1$ is applied then magnetization reversal is not possible with perpendicular magnetic field of any (even infinitely large) amplitude. We call this phenomenon ``rigid saturation''.

The critical current $\varkappa_c=1$ has the following dimensional form
\begin{equation}\label{eq:J0}
J_c=\frac{4\pi M_s^2|e|h}{\hbar\eta}.
\end{equation}
Thus the current $J_c$ is determined only by material parameters (saturation magnetization) and thickness of the film. For the case of permalloy film with thickness $h=20$ nm and rate of spin polarization $\eta=0.4$ the expression \eqref{eq:J0} results $J_c=70.6\times10^{12}\,A/m^2$.

Thus we determine the physical meaning of the value $J_0$ -- the minimal current density which provides the rigid saturation (for the case of full spin-polarization $\eta=1$).

To verify our analytical results we used full-scale OOMMF \cite{OOMMF} micromagnetic simulations. This modelling were performed with material parameters of permalloy as follows: saturation magnetization $M_S=8.6 \times 10^5$ A/m, exchange constant $A=1.3 \times 10^{-11}$ J/m. These values of parameters correspond to the exchange length $\ell=5.3$ nm and saturation field of the infinite film $B_s=1.081$ T. Since the external field and the anisotropy are included into problem in equivalent ways, see Eqs.~\eqref{eq:Ez-lin}, \eqref{eq:Ean}, \eqref{eq:eq-of-motion}, \eqref{eq:crit-current}, in the simulations we restrict ourselves only with case of magnetic field, neglecting the anisotropy ($\alpha=0$). Not being able to simulate he film of infinite size we chose two nanodisks with diameter $D=350$ nm and thicknesses $h=20$ nm and $h=10$ nm and mesh size was $3\times3\times h$ nm. In the absence of the magnetic field and current the ground magnetic state of nanodisks of the mentioned sizes is vortex distribution of magnetization, see Fig.~\ref{fig:Heter}. To these nanodisks we simultaneously apply the external magnetic field of form $B(t)=B_0(1-e^{-t/\Delta t})$ and spin polarized current $J(t)=t\Delta J/\Delta t$, where $\Delta t=1$ ns, $\Delta J=10^{11}\,A/m^2$ with rate of spin polarization $\eta=0.4$. Gradual increase of the field and current allow us to avoid an intense magnon dynamics. Amplitude of the magnetic field was varied in interval $B_0\in[-4,0.95]$~T with step $\Delta B_0=0.05$~T.  The current was increased until the saturation was achieved. As a criterion of saturation we used the relation $M_z/M_s>0.999999$, where $M_z$ is the total magnetization along $\hat{\vec z}$-axis. The resulting dependence $J_s(B)$ in dimensionless form is shown in the Fig.~\ref{fig:B_Jc} by markers. Note a good agreement between theoretical prediction and numerical experiment. The reason for slight discrepancy in region $b\lessapprox1$ is that the saturation field for the finite-size nanodisk $B_s'$ is slightly smaller than $B_s$.

It is known\cite{Volkov11, Gaididei12a} that the VAL usually appear in pre-saturated regime of the ferromagnetic film, see insets of the Fig.~\ref{fig:as_vs_b}. Here we study numerically how the perpendicular magnetic field changes properties of the VAL. We obtained that the positive field (the direction of the field coincides with the current direction) increases the constant of VAL $a_S$ while the negative field (opposite to the current) decreases $a_S$. The resulting dependence $a_S(b)$ is presented in the Fig.~\ref{fig:as_vs_b}. As one can see, the lattice constant $a_S(b)$ is very close to the value $\bar a_S=2\pi/K_0$, where $K_0$ is the wave-vector of unstable magnons for the case $\varkappa\lessapprox\varkappa_s$, see \eqref{eq:crit-current-max}. Assuming that mismatch between $a_S$ and $\bar a_S$ remains small for all values of parameters, one can use \eqref{eq:kappa} to obtain the following asymptotical behavior: $a_S\sim1/(1-b)$ for $b\lessapprox1$ and $a_S\sim1/\sqrt{|b|}$ for $b\rightarrow-\infty$.

\section{Conclusions}
The perpendicular magnetic field drastically changes the process of saturation of magnetic films with spin-polarized current. It is shown that the saturation current $J_s$ is decreased (increased) in case of codirected (oppositely directed) magnetic field and current. There exists a critical current $J_c>J_s$ which provides "rigid" saturation -- the saturated state which is stable with respect to the transverse magnetic field of any amplitude and direction.  The critical current $J_c$ is determined only by material parameters (saturation magnetization) and thickness o the film.  The actions of the perpendicular magnetic field and uniaxial anisotropy on the stability of saturated state are equivalent. The magnetic field changes the constant of the vortex-antivortex lattice $a_S$, which appears in the pre-saturated regime: $a_S$ infinitely increases if the field approaches the saturation value and $a_S$ decreases if the field is increased in the opposite direction. For large opposite fields the fluid-like dynamics of the vortex-antivortex system is observed instead of the static vortex-antivortex lattice.


\begin{thebibliography}{14}
\expandafter\ifx\csname natexlab\endcsname\relax\def\natexlab#1{#1}\fi
\expandafter\ifx\csname bibnamefont\endcsname\relax
  \def\bibnamefont#1{#1}\fi
\expandafter\ifx\csname bibfnamefont\endcsname\relax
  \def\bibfnamefont#1{#1}\fi
\expandafter\ifx\csname citenamefont\endcsname\relax
  \def\citenamefont#1{#1}\fi
\expandafter\ifx\csname url\endcsname\relax
  \def\url#1{\texttt{#1}}\fi
\expandafter\ifx\csname urlprefix\endcsname\relax\def\urlprefix{URL }\fi
\providecommand{\bibinfo}[2]{#2}
\providecommand{\eprint}[2][]{\url{#2}}

\bibitem[{\citenamefont{Lindner}(2010)}]{Lindner10}
\bibinfo{author}{\bibfnamefont{J.}~\bibnamefont{Lindner}},
  \bibinfo{journal}{Superlattices and Microstructures}
  \textbf{\bibinfo{volume}{47}}, \bibinfo{pages}{497} (\bibinfo{year}{2010}).

\bibitem[{\citenamefont{Bohlens et~al.}(2008)\citenamefont{Bohlens, Kr\"{u}ger,
  Drews, Bolte, Meier, and Pfannkuche}}]{Bohlens08}
\bibinfo{author}{\bibfnamefont{S.}~\bibnamefont{Bohlens}},
  \bibinfo{author}{\bibfnamefont{B.}~\bibnamefont{Kr\"{u}ger}},
  \bibinfo{author}{\bibfnamefont{A.}~\bibnamefont{Drews}},
  \bibinfo{author}{\bibfnamefont{M.}~\bibnamefont{Bolte}},
  \bibinfo{author}{\bibfnamefont{G.}~\bibnamefont{Meier}}, \bibnamefont{and}
  \bibinfo{author}{\bibfnamefont{D.}~\bibnamefont{Pfannkuche}},
  \bibinfo{journal}{Appl. Phys. Lett.} \textbf{\bibinfo{volume}{93}},
  \bibinfo{eid}{142508} (pages~\bibinfo{numpages}{3}) (\bibinfo{year}{2008}).

\bibitem[{\citenamefont{Drews et~al.}(2009)\citenamefont{Drews, Kruger, Meier,
  Bohlens, Bocklage, Matsuyama, and Bolte}}]{Drews09}
\bibinfo{author}{\bibfnamefont{A.}~\bibnamefont{Drews}},
  \bibinfo{author}{\bibfnamefont{B.}~\bibnamefont{Kruger}},
  \bibinfo{author}{\bibfnamefont{G.}~\bibnamefont{Meier}},
  \bibinfo{author}{\bibfnamefont{S.}~\bibnamefont{Bohlens}},
  \bibinfo{author}{\bibfnamefont{L.}~\bibnamefont{Bocklage}},
  \bibinfo{author}{\bibfnamefont{T.}~\bibnamefont{Matsuyama}},
  \bibnamefont{and} \bibinfo{author}{\bibfnamefont{M.}~\bibnamefont{Bolte}},
  \bibinfo{journal}{Applied Physics Letters} \textbf{\bibinfo{volume}{94}},
  \bibinfo{eid}{062504} (pages~\bibinfo{numpages}{3}) (\bibinfo{year}{2009}).

\bibitem[{\citenamefont{Kent et~al.}(2004)\citenamefont{Kent, Ozyilmaz, and del
  Barco}}]{Kent04}
\bibinfo{author}{\bibfnamefont{A.~D.} \bibnamefont{Kent}},
  \bibinfo{author}{\bibfnamefont{B.}~\bibnamefont{Ozyilmaz}}, \bibnamefont{and}
  \bibinfo{author}{\bibfnamefont{E.}~\bibnamefont{del Barco}},
  \bibinfo{journal}{Appl. Phys. Lett.} \textbf{\bibinfo{volume}{84}},
  \bibinfo{pages}{3897} (\bibinfo{year}{2004}).

\bibitem[{\citenamefont{Slonczewski}(1996)}]{Slonczewski96}
\bibinfo{author}{\bibfnamefont{J.~C.} \bibnamefont{Slonczewski}},
  \bibinfo{journal}{J.~Magn. Magn. Mater.} \textbf{\bibinfo{volume}{159}},
  \bibinfo{pages}{L1} (\bibinfo{year}{1996}).

\bibitem[{\citenamefont{Berger}(1996)}]{Berger96}
\bibinfo{author}{\bibfnamefont{L.}~\bibnamefont{Berger}},
  \bibinfo{journal}{Phys. Rev. B} \textbf{\bibinfo{volume}{54}},
  \bibinfo{pages}{9353} (\bibinfo{year}{1996}).

\bibitem[{\citenamefont{Slonczewski}(2002)}]{Slonczewski02}
\bibinfo{author}{\bibfnamefont{J.~C.} \bibnamefont{Slonczewski}},
  \bibinfo{journal}{J.~Magn. Magn. Mater.} \textbf{\bibinfo{volume}{247}},
  \bibinfo{pages}{324} (\bibinfo{year}{2002}).

\bibitem[{\citenamefont{Volkov et~al.}(2011)\citenamefont{Volkov, Kravchuk,
  Sheka, and Gaididei}}]{Volkov11}
\bibinfo{author}{\bibfnamefont{O.~M.} \bibnamefont{Volkov}},
  \bibinfo{author}{\bibfnamefont{V.~P.} \bibnamefont{Kravchuk}},
  \bibinfo{author}{\bibfnamefont{D.~D.} \bibnamefont{Sheka}}, \bibnamefont{and}
  \bibinfo{author}{\bibfnamefont{Y.}~\bibnamefont{Gaididei}},
  \bibinfo{journal}{Phys. Rev. B} \textbf{\bibinfo{volume}{84}},
  \bibinfo{pages}{052404} (\bibinfo{year}{2011}).

\bibitem[{\citenamefont{Gaididei et~al.}(2012)\citenamefont{Gaididei, Volkov,
  Kravchuk, and Sheka}}]{Gaididei12a}
\bibinfo{author}{\bibfnamefont{Y.}~\bibnamefont{Gaididei}},
  \bibinfo{author}{\bibfnamefont{O.~M.} \bibnamefont{Volkov}},
  \bibinfo{author}{\bibfnamefont{V.~P.} \bibnamefont{Kravchuk}},
  \bibnamefont{and} \bibinfo{author}{\bibfnamefont{D.~D.} \bibnamefont{Sheka}},
  \bibinfo{journal}{Phys. Rev. B} \textbf{\bibinfo{volume}{86}},
  \bibinfo{pages}{144401} (\bibinfo{year}{2012}).

\bibitem[{\citenamefont{Dussaux et~al.}(2010)\citenamefont{Dussaux, Georges,
  Grollier, Cros, Khvalkovskiy, Fukushima, Konoto, Kubota, Yakushiji, Yuasa
  et~al.}}]{Dussaux10}
\bibinfo{author}{\bibfnamefont{A.}~\bibnamefont{Dussaux}},
  \bibinfo{author}{\bibfnamefont{B.}~\bibnamefont{Georges}},
  \bibinfo{author}{\bibfnamefont{J.}~\bibnamefont{Grollier}},
  \bibinfo{author}{\bibfnamefont{V.}~\bibnamefont{Cros}},
  \bibinfo{author}{\bibfnamefont{A.}~\bibnamefont{Khvalkovskiy}},
  \bibinfo{author}{\bibfnamefont{A.}~\bibnamefont{Fukushima}},
  \bibinfo{author}{\bibfnamefont{M.}~\bibnamefont{Konoto}},
  \bibinfo{author}{\bibfnamefont{H.}~\bibnamefont{Kubota}},
  \bibinfo{author}{\bibfnamefont{K.}~\bibnamefont{Yakushiji}},
  \bibinfo{author}{\bibfnamefont{S.}~\bibnamefont{Yuasa}},
  \bibnamefont{et~al.}, \bibinfo{journal}{Nat Commun}
  \textbf{\bibinfo{volume}{1}}, \bibinfo{pages}{1} (\bibinfo{year}{2010}).

\bibitem[{\citenamefont{Dussaux et~al.}(2012)\citenamefont{Dussaux,
  Khvalkovskiy, Bortolotti, Grollier, Cros, and Fert}}]{Dussaux12}
\bibinfo{author}{\bibfnamefont{A.}~\bibnamefont{Dussaux}},
  \bibinfo{author}{\bibfnamefont{A.~V.} \bibnamefont{Khvalkovskiy}},
  \bibinfo{author}{\bibfnamefont{P.}~\bibnamefont{Bortolotti}},
  \bibinfo{author}{\bibfnamefont{J.}~\bibnamefont{Grollier}},
  \bibinfo{author}{\bibfnamefont{V.}~\bibnamefont{Cros}}, \bibnamefont{and}
  \bibinfo{author}{\bibfnamefont{A.}~\bibnamefont{Fert}},
  \bibinfo{journal}{Phys. Rev. B} \textbf{\bibinfo{volume}{86}},
  \bibinfo{pages}{014402} (\bibinfo{year}{2012}).

\bibitem[{\citenamefont{Sluka et~al.}(2011)\citenamefont{Sluka, K{\'a}kay,
  Deac, B{\"u}rgler, Hertel, and Schneider}}]{Sluka11}
\bibinfo{author}{\bibfnamefont{V.}~\bibnamefont{Sluka}},
  \bibinfo{author}{\bibfnamefont{A.}~\bibnamefont{K{\'a}kay}},
  \bibinfo{author}{\bibfnamefont{A.~M.} \bibnamefont{Deac}},
  \bibinfo{author}{\bibfnamefont{D.~E.} \bibnamefont{B{\"u}rgler}},
  \bibinfo{author}{\bibfnamefont{R.}~\bibnamefont{Hertel}}, \bibnamefont{and}
  \bibinfo{author}{\bibfnamefont{C.~M.} \bibnamefont{Schneider}},
  \bibinfo{journal}{Journal of Physics D: Applied Physics}
  \textbf{\bibinfo{volume}{44}}, \bibinfo{pages}{384002}
  (\bibinfo{year}{2011}).

\bibitem[{\citenamefont{Akhiezer et~al.}(1968)\citenamefont{Akhiezer,
  Bar'yakhtar, and Peletminski\u{\i}}}]{Akhiezer68}
\bibinfo{author}{\bibfnamefont{A.~I.} \bibnamefont{Akhiezer}},
  \bibinfo{author}{\bibfnamefont{V.~G.} \bibnamefont{Bar'yakhtar}},
  \bibnamefont{and} \bibinfo{author}{\bibfnamefont{S.~V.}
  \bibnamefont{Peletminski\u{\i}}}, \emph{\bibinfo{title}{Spin waves}}
  (\bibinfo{publisher}{North--Holland}, \bibinfo{address}{Amsterdam},
  \bibinfo{year}{1968}).

\bibitem[{OOM()}]{OOMMF}
\emph{\bibinfo{title}{The {O}bject {O}riented {M}icro{M}agnetic {F}ramework}},
  \bibinfo{note}{developed by M. J. Donahue and D. Porter mainly, from NIST. We
  used the 3D version of the 1.2$\alpha$4 release},
  \urlprefix\url{http://math.nist.gov/oommf/}.

\end{thebibliography}

\end{document}